\documentclass[fleqn,usenatbib]{mnras}
\usepackage{multicol}
\usepackage[T1]{fontenc}
\usepackage{ae,aecompl}
\usepackage{graphicx}	
\usepackage{amsmath}	
\usepackage{amssymb}	
\usepackage{multirow}
\title[Long-term variability of RRATs]{Pushchino multibeam pulsar search - II. Long-term variability of RRATs.}

\author[T. V. Smirnova et al.]{
T. V. Smirnova$^{1}$\thanks{E-mail: serg@prao.ru},
S. A. Tyul'bashev$^{1}$,
E. A. Brylyakova$^{1,2}$,
M. A. Kitaeva$^{1}$,
I. V. Chashei$^{1}$, 
\newauthor{
G. E. Tyul'basheva$^{3}$,
V. V. Oreshko$^{1}$, 
S. V. Logvinenko$^{1}$
}
\\ 
$^{1}$ P.N. Lebedev Physical Institute of the Russian Academy of Sciences, Astro Space Center,\\ 
Pushchino Radio Astronomy Observatory\\
Radiotelescopnaya 1a, Moscow reg., Pushchino, 142290, Russia \\
$^{2}$ P.N. Lebedev Physical Institute of the Russian Academy of Sciences\\
Leninskii pr. 53, Moscow, 119991  Russia \\
$^{3}$ Institute of Mathematical Problems of Biology, brunch of Keldysh Institute of Applied Mathematics \\
Vitkevich 1, Moscow reg., Pushchino, 142290, Russia
}

\date{Accepted XXX. Received YYY; in original form ZZZ}
\pubyear{2021}
\begin{document}
\label{firstpage}
\maketitle
\begin{abstract} 
Pulses from 16 previously known rotating radio transients (RRAT) have been searched at the 110~MHz daily monitor program for 4 to 5.5 years by using the Large-Phased-Array (LPA) at Pushchino. The total number of pulses detected such a long observation interval is only 90 pulses for RRAT J0640+07 or is as large as 10751 pulses for RRAT J0302+22. The number and amplitude of pulses varies at a time-scales from six to twenty months for RRATs J1336+33, J1404+11, J1848+15, J2051+12, J2105+22, and the pulse number can increase by one or two orders of magnitude in active phases. The long-term trends are found for RRATs J0139+33 and J0302+22, showing a 2-3 times increase in detected pulse number over 1959 days. Some RRATs show the annual variations on both pulse number and pulse amplitude. It is hard to explain all these variation time scales by refractive scintillation on the interstellar medium. The annual and semi-annual variations are likely caused by scintillations of the inhomogeneous interplanetary plasma. Our data show that the number of observational sessions with no pulse detection over the threshold decreases exponentially with the length of pulse silence.
\end{abstract}

\section{Introduction}

Despite the 15 years that have passed since the discovery of RRATs, there is neither a complete understanding of what these objects are, nor even a good definition of a RRAT. A set of properties is known that indicate that the object under study can be RRAT: the object is detected by individual dispersed pulses; it has a typical period derivatives and periods larger than that of normal (ordinary) pulsars ($P>30$~ms); the time between detected pulses can be from tens of seconds to tens of hours; the mean value of magnetic field of RRAT is the larger than the magnetic field of ordinary pulsars; on a $P/\dot P$ diagram RRATs are often located near the ``death line'' or in the transition zone between ordinary pulsars and magnetars; their single-pulse amplitude distributions are similar with pulsars: log-normal, log-normal with power-law tail, power law distribution; the Galactic z-distribution of RRATs is similar with pulsar z-distribution; RRAT pulse widths are similar with width of pulsars of similar periods (\citeauthor{McLaughlin2006}, \citeyear{McLaughlin2006}, \citeauthor{Burke-Spolaor2009}, \citeyear{Burke-Spolaor2009}, \citeauthor{Keane2010}, \citeyear{Keane2010}, \citeauthor{Karako-Argaman2015}, \citeyear{Karako-Argaman2015},
\citeauthor{Cui2017}, \citeyear{Cui2017}, \citeauthor{Logvinenko2020}, \citeyear{Logvinenko2020}, \citeauthor{Tyulbashev2021}, \citeyear{Tyulbashev2021}). Not all the noted properties are observed for each RRAT, and therefore it is not clear whether it is possible to give a strict definition for a RRAT, or the term ``rotating transients'' is a conditional name for pulsars with special properties.

For some of the pulsars detected as RRATs, their periodic emission was later found. Some pulsars detected as RRATs at high frequencies look like ordinary pulsars, albeit with individual strong pulses in the meter wavelength range.

The variability of pulsars radio emission is detected at very different time scales. This variability can be either intrinsic, associated with the pulsar itself, or external, caused by scintillations on the interstellar plasma. The manifestations of proper variability can be different: a presence of a microstructure in a number of pulsars with scales from microseconds to milliseconds, an absence of emission from one to many periods (nullings), a sharp change in a shape of the pulse: a mode switching  (\citeauthor{Backer1970b}, \citeyear{Backer1970b}, \citeauthor{Backer1970a}, \citeyear{Backer1970a}), giant pulses with a duration of up to nanoseconds  (\citeauthor{Hankins2003}, \citeyear{Hankins2003}). The time scale of mode and nullings changes can be from several periods to several hours or even days (\citeauthor{Wang2007}, \citeyear{Wang2007}). There are pulsars that have pseudo periodic switching off and on of emission (intermittent pulsars). For example, the pulsar B1931+24 behaves this way for long time intervals with periods of the emission switching ``off'' and turning ``on'' with the scales of 25-35 days and 5-10 days, respectively (\citeauthor{Kramer2006}, \citeyear{Kramer2006}). 

The external variability is associated with the passage of emission from a pulsar through the inhomogeneities of the interstellar plasma, which causes a number of observed effects: an increase in the angular size of a source, a change in a shape of a pulse, temporal intensity and frequency variations (scintillation). There are diffraction and refractive scintillation (\citeauthor{Rickett1990}, \citeyear{Rickett1990}). The typical time scale of diffraction scintillation on meter wavelength is seconds and tens of seconds (\citeauthor{Smirnova1992}, \citeyear{Smirnova1992}), the typical time scale of refractive scintillation is months and years. 

One of the extreme manifestations of the disappearance of emission is observed in a special type of pulsars called rotating radio transients (RRAT). RRATs were discovered by \citeauthor{McLaughlin2006} (\citeyear{McLaughlin2006})  as individual dispersed pulses. The appearance of pulses is irregular in time. The pulses of RRATs are usually weak and therefore, when searching for RRAT, the fluctuation sensitivity of the radio telescope is the determining factor for their detection. The study of already detected RRATs is difficult. The average time intervals between the recorded RRAT pulses can be from tens of seconds and minutes and reach up to several hours \citeauthor{McLaughlin2006} (\citeyear{McLaughlin2006}) or even tens of hours (\citeauthor{Logvinenko2020}, \citeyear{Logvinenko2020}). Since the appearance of pulses is sporadic, tens, hundreds or even thousands of hours of observations of a RRAT are needed for investigation of RRATs.

There have just a few studies of the long-term variability of RRATs (\citeauthor{Palliyaguru2011}, \citeyear{Palliyaguru2011}, \citeauthor{Bhattacharyya2018}, \citeyear{Bhattacharyya2018}). In the paper of \citeauthor{Palliyaguru2011} (\citeyear{Palliyaguru2011}), an analysis of the short- and long-term variability of eight RRATs was done, studied at the interval of 5.5 years. A noticeable increase in the rate of pulses was registered for 6 of them at some time intervals. In the paper of \citeauthor{Bhattacharyya2018} (\citeyear{Bhattacharyya2018}), on over an eight-year observation interval for three RRATs, it was shown that there are trends in increasing or decreasing the observed number of pulses over time. For one of three RRATs, bursts of emission were shown when the number of recorded pulses per hour increased tenfold and more times from one epoch to another epoch.

In 2017 when processing semi-annual daily observations obtained on the Large Phased Array (LPA) radio telescope of P.N.Lebedev Physics Institute (LPI), 33 transients were discovered by their individual dispersed pulses (\citeauthor{Tyulbashev2018b}, \citeyear{Tyulbashev2018b}, \citeauthor{Tyulbashev2018a}, \citeyear{Tyulbashev2018a}). When processing the data, individual pulses from more than 100 known pulsars were also detected\footnote{http://bsa-analytics.prao.ru/en/transients/pulsars/}. Some of these pulsars are in the RRATalog catalog\footnote{http://astro.phys.wvu.edu/rratalog} and are considered to be RRATs.  Monitoring observations on the LPA LPI antenna have been going on for more than five years.  Based on the processing of monitoring observations, a cycle of work has begun on the search and study of pulsars of different types. In the first paper of the cycle  (Tyul'bashev et al., submitted) a method for searching for pulsars with an expected integral flux density less than 0.2 mJy at the frequency 111 MHz is considered. The goal of this paper is to study of the variability of the RRAT pulsars discovered by the LPA LPI.

\section{Observations and calibration}

Monitoring observations by the meridian antenna of the LPA LPI were carried out in the 2.5 MHz band with the central frequency of 110.3 MHz. The band was divided into 32 frequency channels with a channel width of 78 kHz. The sampling time is 12.5 ms. The observation session used for data processing had a duration of 3 minutes, which approximately corresponds to the time of the source passing through the center of the antenna beam at the level of 0.5.

Round-the-clock daily monitoring PUshchino Multibeam Pulsar Search (PUMPS) started in mid-August 2014. It is going on simultaneously in 96 antenna beams, overlapping declination angles from $-9^o$ to $+42^o$. Six times a day, a signal of a known temperature is sent to the antenna input in the form of ``OFF-ON-OFF''  (\citeauthor{Tyulbashev2019}, \citeyear{Tyulbashev2019}). This allows to equalize the gain in the frequency channel. The change in the height of the calibration signal does not exceed 5\%. Since the closest calibration signal to the source is used for
session calibration, the detected pulse amplitudes have errors
of less than 5\%. Sessions with interference were excluded from our data analysis. Additional information about the observation regimes of the LPA LPI, is given in the papers \citeauthor{Tyulbashev2016} (\citeyear{Tyulbashev2016}, \citeauthor{Shishov2016} (\citeyear{Shishov2016}). 

The selection of sources for this work was made based on the fact that the number of detected pulses should be sufficient to study long-term variability, and the rotating transient itself meets one of the following criteria: a) the object was detected as RRAT in our observations and has not yet been confirmed by other authors; b) the object was detected as an RRAT at a frequency of 111 MHz and as an ordinary pulsar in the observations of other authors; c) the object was detected as pulsar in observations at a frequency of 111 MHz, but is in the RRATs lists according to the observations of other authors. In total, there were 15 objects in the list of studied sources, which we detected by individual pulses.

The pulsar J0659+14 (B0656+14) was added to this list, which, as the authors of the paper \citeauthor{Weltevrede2006} (\citeyear{Weltevrede2006})  suggest, would look like an object of the RRAT type if it was placed on for a large distance. The pulsar has a long tail in the distribution of pulses over amplitudes. For pulsars with such a distribution, regular (periodic) emission may not be detected due to insufficient sensitivity in the surveys, but the strongest pulses from the tail of the distribution may be observed. In search observations by the LPA LPI, the source is detected both in individual pulses and as an ordinary pulsar.

The first column of Table 1 shows the name of the source, the columns 2-3 - galactic longitude and latitude.  The estimates in columns 4-6 are taken from the work \citeauthor{Sanidas2019} (\citeyear{Sanidas2019}), if the sources were observed on LOFAR at a frequency of 135 MHz, or our estimates are given from the works (\citeauthor{Tyulbashev2018b}, \citeyear{Tyulbashev2018b}, \citeauthor{Tyulbashev2018a}, \citeyear{Tyulbashev2018a}). The periods of the sources J1005+30 and J2105+19 are indicated in \textbf{bold}, the estimates for which are obtained in this paper. The periods were obtained as the greatest common divisor for sessions in which two or more pulses were observed. The actual period can be less by an integer number of times. The periods of RRATs J0139+33, J1132+25, J1336+33, J1502+28 were estimated in the same way in our early work 
(\citeauthor{Tyulbashev2018a}, \citeyear{Tyulbashev2018a}).

Unfortunately, we could not get timing solution for our sources. The accuracy of the timestamps are determined by the quartz oscillator ($\pm 25$~ms at laboratory conditions). Additionally we have bad coordinates accuracy (typical $\pm 1^m$ in right ascension and $\pm 15^\prime$ in declination). In a result we have a lot of solution with possible periods and coordinates of sources and cannot choose best solution.

\begin{table}
\caption{The list of the investigated sources and some of their characteristics: galactic longitude and latitude, period ($P$), dispersion measure ($DM$), pulse width at $50\%$ of peak  ($W_{0.5}$). The superscripts 'a', 'b', 'c' used for each source correspond to a critaria  described in the text of Section~2.}
\begin{tabular}{lccccclllllll}
\hline

Name & l, & b, & P, & DM,  & $W_{0.5}$,  \\
     & deg.& deg. & s & pc/cm$^3$ & ms \\
\hline
J0139+331$^a$ & 134.38 & -28.17 &1.2479 & 21.2        & 25      \\
J0302+223$^c$ & 158.44 & -30.84 &1.2072 & 19.09       & 48      \\
J0317+132$^b$ & 168.75 & -36.03 &1.9743 & 12.9        & 20      \\
J0609+161$^a$ & 193.42 & -1.504 &0.9458 & 85          & 55      \\
J0640+071$^a$ & 204.88 & 1.252 &       & 52          & 35      \\
J0659+14  &     201.11 & 8.258 &0.3849 & 13.9        & 15      \\
J1005+301$^a$ & 197.94 & -28.17 & {\bf 3.069}  & 17.5        & 30      \\
J1132+251$^a$ & 214.58 & 72.28 & 1.002  & 23          & 20      \\
J1336+331$^a$ & 70.32 & 78.34 & 3.013  & 8           & 15      \\
J1404+112$^b$ & 355.02 & 67.10 & 2.6505 & 18.4        & 53      \\
J1502+281$^a$ & 42.78 & 61.13 & 3.784  & 14          & 25      \\
J1538+233$^c$ & 37.32 & 52.39 & 3.4494 & 14.9        & 25      \\
J1848+152$^b$ & 46.33 & 7.45 & 2.2338 & 77.4        & 45      \\
J2051+122$^b$ & 59.35 & -19.44 & 0.5532 & 43.4        & 50      \\
J2105+191$^a$ & 66.998 & -18.18 & {\bf 3.5292} & 33          & 25      \\
J2209+222$^b$ & 79.92 & -27.78 & 1.7769 & 46.3        & 36 \\ 
\hline
\end{tabular}
\\
Notes: In observations by LOFAR (\citeauthor{Sanidas2019}, \citeyear{Sanidas2019}) RRATs J0302+22, J0317+13, J1404+11, J1848+15 were initially found by individual pulses, but then their periodic emission, characteristic of ordinary pulsars, was also shown. J2051+12 in search at 111 MHz was detected by single pulse search (\citeauthor{Tyulbashev2018a}, \citeyear{Tyulbashev2018a}), and in the search by LOFAR (\citeauthor{Sanidas2019}, \citeyear{Sanidas2019}), it was detected as a regular pulsar. The transient J2209+22 was detected using LPA LPI first 
(\citeauthor{Tyulbashev2018a}, \citeyear{Tyulbashev2018a}), but later its periodic emission was detected (\citeauthor{Sanidas2019} (\citeyear{Sanidas2019}), \citeauthor{Tyulbashev2020} (\citeyear{Tyulbashev2020})). The transients J0301+20 in RRATalog (J0302+22 in our observations) and J1538+23 in the paper \citeauthor{Karako-Argaman2015} (\citeyear{Karako-Argaman2015}) were determined as RRAT. In ATNF catalog\footnote{https://www.atnf.csiro.au/people/pulsar/psrcat/}, they are listed as pulsars. In our search for transients, both periodic emission and individual pulses of these sources were detected.
\end{table}

There are 16 sources in Table 1. One of them (J0659+14) is known as an ordinary pulsar with an unusually long tail in the distribution of pulses by energy. RRAT J1538+23 discovered by \citeauthor{Karako-Argaman2015} (\citeyear{Karako-Argaman2015}) is ordinary pulsar at 111 MHz. The remaining 14 sources discovered in \citeauthor{Tyulbashev2018b} (\citeyear{Tyulbashev2018b}), 
\citeauthor{Tyulbashev2018a} (\citeyear{Tyulbashev2018a}) as RRATs. For eight sources (J0302+22; J0317+13; J0609+16; J1404+11; J1538+23; J1848+15; J2051+12; J2209+22) the ordinary pulsar emission is also detected. Therefore, the question arises whether there is any difference between these sources and ordinary pulsars detected when searching for RRATs by their individual pulses. Seven sources (J0139+33; J0302+22; J0317+13; J1404+11; J1848+15; J2051+12; J2209+22) were re detected by LOFAR as ordinary pulsars or as RRAT. For these sources LOFAR gives average profile, dispersion measure and period better than LPA, but they didn't study the variability of strong single pulses ($S/N \ge 6$), emission rates and there behaviour over long time as we do.

For all sources from Table 1, the search for pulses was carried out over a time interval up to 5.5 years. The search for RRAT pulses in the survey data was carried out according to the standard scheme: gain equalization between frequency channels; compensation of a known dispersion measure of the studied transient; subtracting of the baseline; detection of signals having $S/N \ge 6$. We did monitoring of radio frequency interference and through away data with it. Just a few percents of data were excluded for this reason. The detected signals, graphics of the pulse profile and dynamic spectrum were saved, and the final selection of pulses was made visually. The details of the processing can be found in the paper \citeauthor{Brylyakova2021} (\citeyear{Brylyakova2021}).

\section{Analysis of pulse emission}

\subsection{Characteristics of pulse emission}

The general characteristics of pulse emission for 16 studied sources obtained over a time interval of 4-5.5 years are shown in Table~2. In the column 2-7, there are: a date of the first pulse detection; total number of observation days, $N_{all}$; a number of recorded pulses, $N_{puls}$; the ratio of the number of days in which the pulses were detected to the total number of observation days, $f_{obs}$ (intermittency factor); the average number of pulses per three-minute observation session, $N_p$ ($N_p$ was determined only by sessions in which at least one pulse was detected); the maximum continuous duration of the absence of pulses exceeding threshold ($S/N \ge 6$) in days (no pulses in each session) for the entire observation interval, $N_{0,max}$.  Columns 8-10 show the average values of amplitudes in signal-to-noise ratio units ($S/N$) for all detected pulses, exceeding the specified threshold, $\langle I \rangle _1$; the average value taking into account the intermittency factor, $\langle I \rangle _2 = \langle I \rangle _1 \times f_{obs}$; the maximum value of $S/N$ for the entire observation period, $I_{max}$. In column 11, there is modulation index $m_1$: 

\begin{equation}
m_1 = \sum\limits_{i=1}^N [(I_i - \langle I \rangle_1)^2 /(N - 1)]^{1/2}/ \langle I \rangle_1 ,
\end{equation}

where $N$ is the number of pulses. The modulation index $m_1$ was determined by the variations in the amplitudes (in $S/N$ units) for all individual pulses over the entire time interval. 

\begin{table*}
	\centering
	\caption{Characteristics of the detected pulses}
	\begin{tabular}{cccccccccccc}
			\hline
Name & Date & $N_{all}$ (days) & $N_{puls}$ & $f_{obs}$  & $N_{p}$ & $N_{0,max}$ (days)  & $\langle I \rangle_1$ & $\langle I \rangle_2$ & $I_{max}$ & $m_{1}$ & $t_1$ (days)\\
	\hline
J0139+33&	21.08.2014&	1959&	2074&	0.62&	1.71&	29&	20.1&	12.1&	218&	0.95 & {0.93 $\pm$ 0.02}\\
J0302+22&	21.08.2014&	1959&	10751&	0.81&	6.77&	9&	11.5&	9.35&	127.5&	0.52 & {0.59 $\pm$ 0.01}\\
J0317+13&	03.01.2015&	1501&	1798&	0.56&	2.14&	34&	9&	5.0&	33.4&	0.33 & {1.14 $\pm$ 0.06}\\
J0609+16&	09.01.2015&	1516&	250&	0.107&	1.54&	59&	8.7&	0.93&	23.8&	0.34 & {5.76 $\pm$ 0.57}\\
J0640+07&	19.09.2014&	1927&	90&	0.044&	1.05&	107&	14&	0.62&	289&	2.2 & {11.7 $\pm$ 2.8} \\
J0659+14&	22.08.2014&	1955&	864&	0.44&	1.4&	20&	11.3&	4.97&	71.8&	0.57  & {2.5 $\pm$ 0.1} \\
J1005+30&	28.08.2014&	1942&	389&	0.19&	1.06&	31&	38.3&	7.2&	448&	1.25 & {4.2 $\pm$ 0.3} \\
J1132+25&	26.08.2014&	1937&	206&	0.075&	1.42&	265&	7.13&	0.53&	16.2&	0.2 & {1.45 $\pm$ 0.18}\\
J1336+33&	26.08.2014&	1896&	157&	0.069&	1.21&	256&	10.4&	0.71&	41.3&	0.5 & {2.9 $\pm$ 0.4}\\
J1404+11&	02.01.2015&	1822&	3492&	0.47&	4.05&	31&	10&	4.7&	81.6&	0.34 & {1.68 $\pm$ 0.06}\\
J1502+28&	22.08.2014&	1603&	4363&	0.54&	5.06&	9&	7.3&	3.94&	73.8&	0.38 & {1.31 $\pm$ 0.04} \\
J1538+23&	21.08.2014&	1956&	2153&	0.49&	2.26&	35&	10.9&	5.3&	61.3&	0.45 & {1.51 $\pm$ 0.06} \\
J1848+15&	22.08.2014&	1494&	307&	0.16&	1.31&	55&	11.2&	1.8&	34.9&	0.31 & {3.56 $\pm$ 0.38}\\
J2051+12&	27.08.2014&	1951&	265&	0.12&	1.15&	49&	9.5&	1.14&	42.7&	0.31 & {8.1 $\pm$ 0.8}\\
J2105+19&	21.08.2014&	1951&	1327&	0.29&	2.34&	30&	9&	2.6&	32.8&	0.28 & {2.07 $\pm$ 0.09} \\
J2209+22&	23.08.2014&	1816&	260&	0.102&	1.41&	126&	7.25&	0.74&	15&	0.21 & {5.4 $\pm$ 0.6} \\
\hline
\end{tabular}
\end{table*}

From the Table~2 it can be seen that the parameters of different sources differ significantly. The recorded number of pulses for the entire observation period ($N_{puls}$)  changes from 90 (J0640+07) to 10751 (J0302+22). Maximum amplitude of detected pulses ($I_{max}$), expressed in $S/N$ ratio units, changes from 15 (J2209+22) to 448 (J1005+30), modulation index $m_1$ changes from 0.2 (J1132+25) to 2.2 (J0640+07). Average number of detected pulses per observation session ($N_p$) for cases when at least one pulse was detected during the session (3 minutes per day) changes from 1 (J0640+07) to 6.8 (J0302+22). 

It can be noted from the Table~2 that half of the transients have a small value of the average number of pulses per observation session (3 min), $N_p \le 1.5$. Usually it is also accompanied by a small intermittency factor, $f_{obs} \le 0.2$. Generally (except for sources J0139+33, J0317+13, J0659+14, J1538+33) there is a proportional dependence of the intermittency factor vs. $N_p$. The largest continuous number of sessions with no pulses having $S/N > 6$ (nullings, parameter $N_{0,max}$), was detected for J1132+25 and J1336+33 (265 and 256 days), that was noted for these RRATs (\citeauthor{Tyulbashev2021}, \citeyear{Tyulbashev2021}). For two more sources, $N_{0,max}$ exceeds 100: J0640+07 (107 sessions), J2209+22 (126 sessions). These two pulsars have also small values of $N_p$ and $f_{obs}$. In general, the number of days with no pulses has a fairly wide distribution from one session to the maximum length given in Table 2, column 7. We will analyze these distributions later. The highest modulation indices are observed for pulsars with the highest values of $I_{max}$ (J0640+07, J1005+30). The longest duration of the absence of pulses $N_{0,max}$ is realized for transients, for which few pulses per session are recorded. The shortest duration of $N_{0,max}$ is observed for sources with the highest value of $N_p$. There is no obvious connection between the nulling duration and the modulation index $m_1$. The properties of the normal pulsar J0659+14 do not differ strongly from another sources from our list.

\begin{figure}
	\includegraphics[width=0.99\columnwidth]{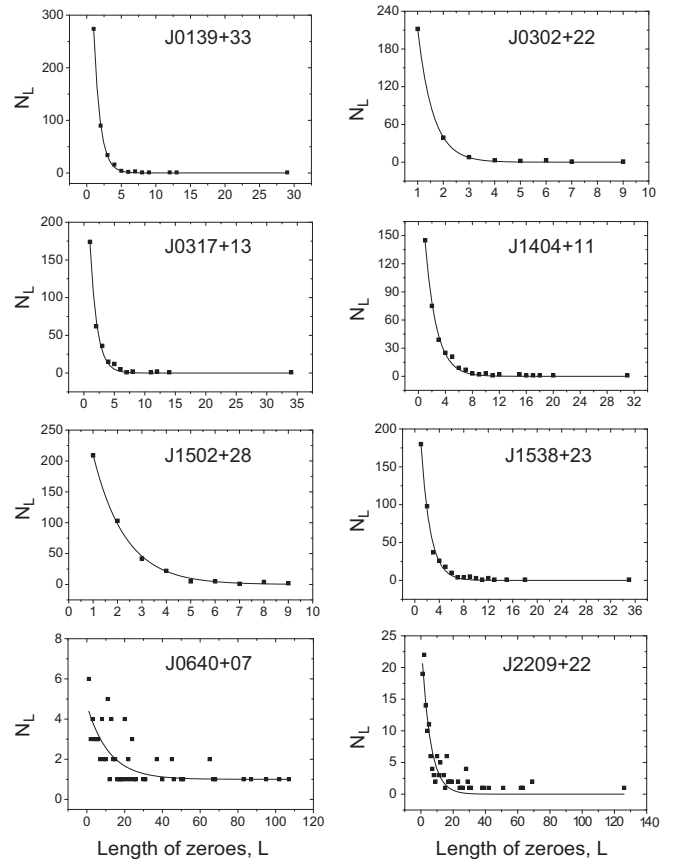}
	\caption{Dependence of a number of realizations, $N_L$, vs. a number of sessions with the absence of pulses in them with an amplitude higher than the detection threshold (nulling length), $L$. The line corresponds to the fitting of the exponential function using the least squares method.}
\end{figure}

We examine here the distribution of nullings. An array was created for each source, in which the number of elements is equal to the number of observation days. If at least one pulse was observed during the session, the value ``1'' was assigned to this day (element), if no pulse was observed, the value of the element was assumed to be ``0''. Fig.~1 shows examples of the distribution of the number of cases (sessions) of the absence of pulses, $N_L$, depending on their duration (a continuous number of days without pulses), L, for 8 sources. Although the session duration was only 3 minutes, statistically they will reflect the real distributions. The probability of no pulses in one session when recording 1 pulse in 3 minutes is 50\%, and it will, accordingly, decrease with increasing L. Into the resulting histograms, an exponential function was fitted using the least squares method:

\begin{equation}
N_L = a \times e^{-L/t_1}
\end{equation}

The parameter $t_1$ with errors for all sources is shown in the last column of the Table~2. The second column in the table shows the time scale of the nulling length (the absence of pulses exceding $S/N\ge 6$), at which the number of events, $N_L$, falls by ‘e’ times. The two lower diagrams in Fig.~1 for sources J0640+07 and J2209+22 show examples of a strong spread of histogram points from the approximation by an exponent. For J0640+07 and for J2209+22, the small number of pulses was detected, $N_{puls}$ = 90 and 260.

For 12 sources for which there are long nullings ($L> 20$), the number of realizations $N_L$ does not exceed two, however, the length of the nullings for these single realizations varies widely: from 20 to 126. Two RRATs (J2051+12, J2209+22) have an exceedance of $N_L$: $N_L = 4$ for $L = 28$ and $N_L = 6$ for $L = 22$, respectively. J1502+28 and J0302+22 have the smallest range of nulling lengths: $L$ from $1$ to $9$. The largest number of registered pulses and the largest average number of pulses per one session are realized for them. $t_1$ characterizes the degree of intermittency: than larger $t_1$ than less intermittency factor $f_{obs}$ (for $t_1 \gtrsim 2.9$, $f_{obs}<0.2$).

\subsection{Long-term variability}

In order to consider the variability in the interval of a month or more, we presented our data in Fig.2 row by row: the first - the number of pulses detected per month, the second - the average amplitude in S/N ratio units of all detected pulses for each month. The interval of one month was chosen for the reasons that for an object with a minimum number of detected pulses, it would be possible to visually observe the dependence, if there is one. The minimum number of detected pulses is equal to 90 for J0640+07. Since about 60 months were processed, we can expect an average of 1.5 events per month for J0640+07.

The sources studied by us have a strong variability, both for short (months) and long (years) time intervals. Fig.~2 shows that the number of pulses observed per month can change dramatically (up to 39 times) in short time intervals (1-2 months: J0609+16, J1132+25, J2209+22) and in the intervals within the order of 6-20 months up to 50 times (J0317+13, J1336+33, J1404+11, 1848+15, J2051+12, J2105+22). The largest change in the number of detected pulses by 50 times on the interval of 6 months is observed for J1404+11. In addition to variability on a scale of several months, for RRATs J0139+33, J0302+22 long-term trends are clearly visible. For these sources, Fig.~2 shows the approximation of the trend by the straight line ($y = a + bx$) using the method of least squares. The following coefficients were obtained: $a=22.8 \pm 2.2$, $b=0.28 \pm 0.06$ (for J0139+33) and $a=119.0 \pm 9.3$, $b=1.44 \pm 0.25$ (for J0302+22). Long term variability observed for J0139+33 and J0302+22, has increase of pulse emission rate of 0.28 pulse/month and 1.44 pulse/month respectively. These trends show that the observed number of pulses can generally increase by 2-3 times over a five-year interval. It was previously noted in the paper \citeauthor{Bhattacharyya2018} (\citeyear{Bhattacharyya2018}). 

\begin{figure*}
	\includegraphics[height=0.9\textheight]{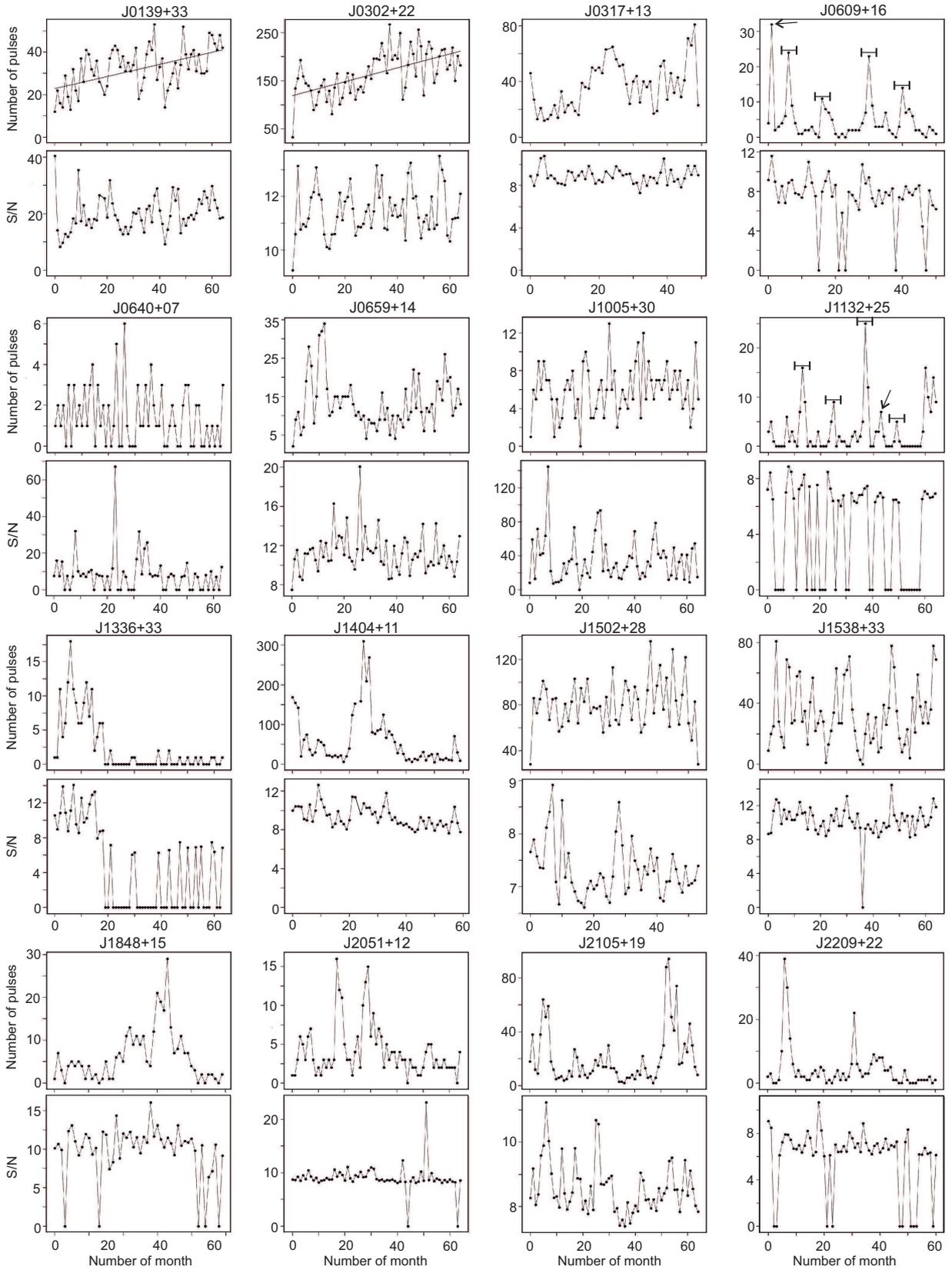} 
	\caption{The dependences of the total number of pulses and S/N recorded for each month of observations vs. time are presented row by row.  Along the x-axis, there is time in months, along the y-axis, there are the number of detected pulses and S/N. The values are indicated by dots. The points are connected by a line to make the character of the changes more visible. The points with zero level correspond to months with pulses absence. For J0139+33, J0302+22, the straight line shows long-term trends. The segments and arrows plotted on the pictures for some RRATs are explained in section 3.4.}
\end{figure*}

\begin{figure}
	\includegraphics[width=\columnwidth]{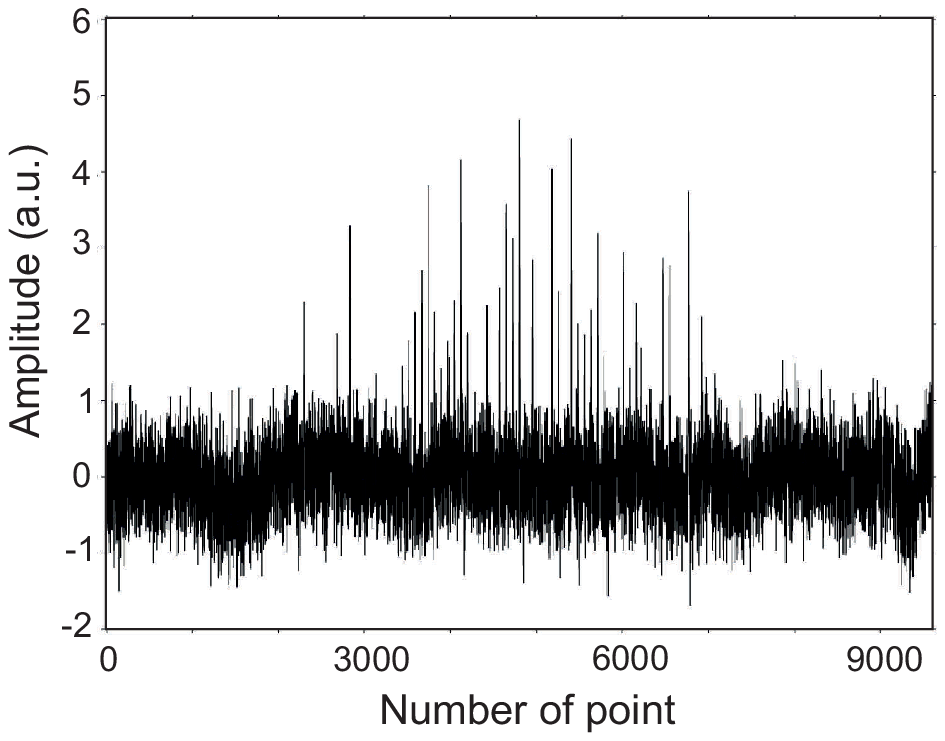}
	\caption{\textbf{The real data recording for J0609+16 on February 17 2015. There are 9600 data points over a daily two-minute session, with the sample rate of $\Delta t = 12.5$~ms. Some strong pulses are detected in the range of 2500-7000.}}
\end{figure}

A very interesting example of variability was demonstrated by source J0609+16. During its study, it turned out that the first strong peak, visible in Fig.~2 (see arrow), was formed during one day of observations. It turned out that on February 17, 2015, 32 pulses with $S/N \ge 6$ were registered within 57.7 seconds. In the data presented in Fig.~3 in two minutes of recording, it can be seen that in addition to 32 strong pulses, six weaker pulses $(S/N<6)$ were also observed in the recording, which were not detected in the search. In total, visually 38 pulses were detected in this session. For the whole of February, there was only one more day (February 24, 2015), when 2 pulses with a $S/N \ge 6$ were registered. This reflects a burst character of emission for  this source: only 32 pulses were detected for 6.5 months from March 2015. 

Fig.~2 shows the change in the average pulse amplitude for each month (in S/N ratio units) over time. Its values, shown by dots, are connected by lines, although no one pulse above the specified threshold was registered in some months (with points we placed at zero level). Some of peaks seen in number of pulses coincide with the peaks in S/N, but in general there is no correlation. The burst activity in terms of the number of pulses is not accompanied by an increase in the signal amplitude. We also note that clearly expressed in Fig.~2 long-term trends of increasing the number of observed pulses for transients J0139+33 and J0302+22 are absent in $S/N$ plots. A sharp change in the signal amplitude at an interval of 1 month (up to 10 times) is observed for J0139+33, J0640+07, J1005+30, J2051+12. The transient J1336+33 has burst activity both in the number of pulses and in the amplitude for about 20 months, and then for 40 months only single rare and weak pulses were recorded. Such a change in the activity of J1336+33 was previously noted in the paper \citeauthor{Brylyakova2021} (\citeyear{Brylyakova2021}).


Another way to find periodic processes is to use autocorrelation functions (ACF) for temporal series analysis. Sometimes the detected periodicities in the ACF look clearer. We performed a correlation analysis of the same arrays that we used for spectral analysis. Fig.~4 shows the normalized correlation functions for several sources. The values for zero shift are removed. The figure clearly shows the periodicity for time shifts corresponding to one year.

In addition to the annual periodicity, Fig.~4 also shows a weakly expressed semi-annual periodicity. The annual periodicity from the ACF analysis was found for 5 sources: J0609+16, J1132+25, J2051+12, J2105+19, J2209+22. Plots with ACF for all sources are given on our website\footnote{http://prao.ru/online\%20data/onlinedata.html}.



\begin{figure}
	\includegraphics[width=\columnwidth]{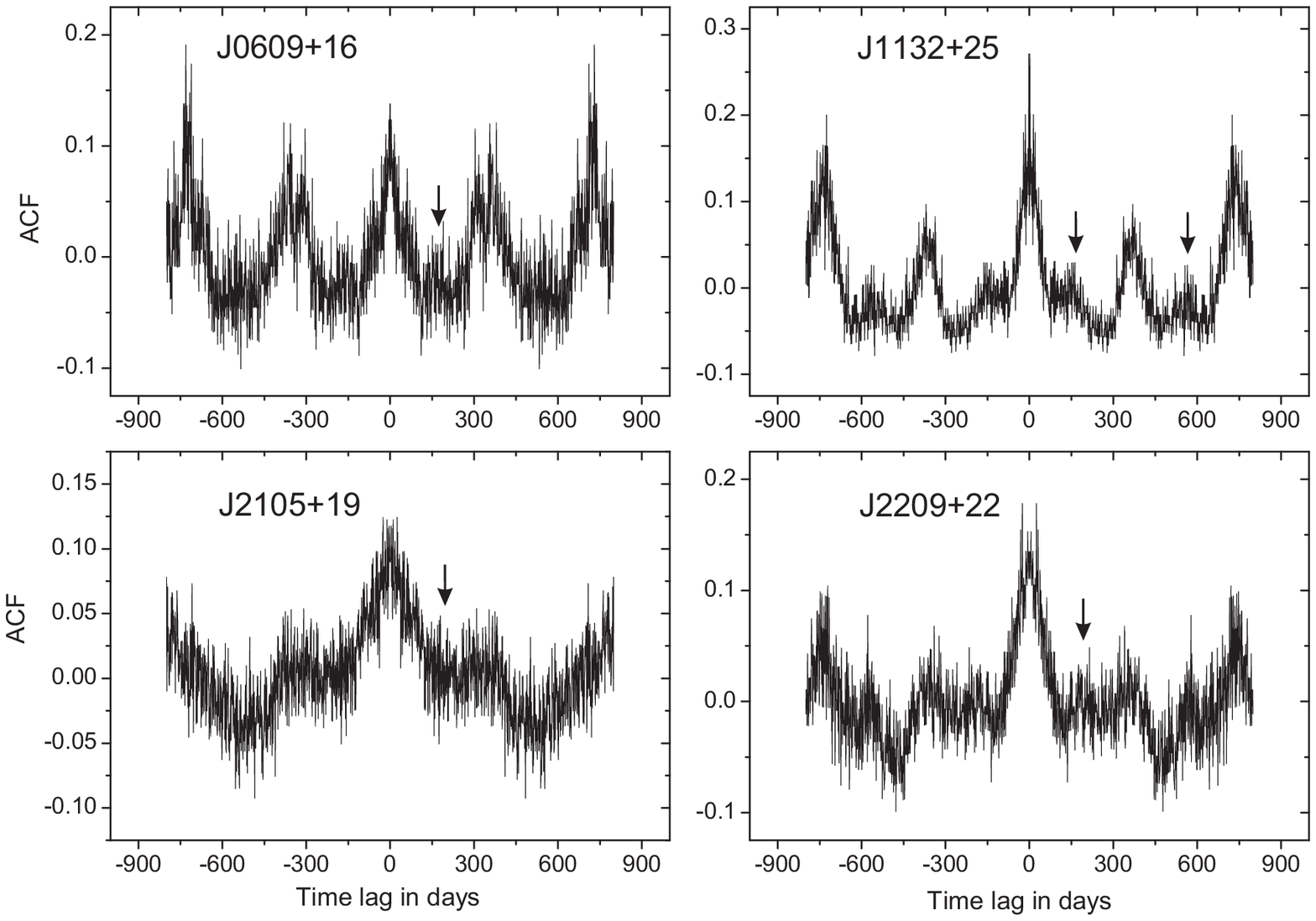}
	\caption{Normalized autocorrelation function from arrays of zeros (no pulses) and ones (days when there is at least one pulse) for four sources. The horizontal axis shows shifts in days, the vertical axis shows the amplitude of ACF. The arrows indicate the probable semi-annual periodicities.}
\end{figure}

\subsection{Effects of interstellar scintillation in amplitude variations}


As we have already noted above, the variations in the amplitude and number of pulses over time can be determined both by the internal variability of the sources and by the passage of emission through the inhomogeneities of the interstellar plasma. These inhomogeneities cause the scattering of radio waves, which leads to variations in the signal amplitude, both in frequency and in time, it is a diffraction interstellar scintillation. In addition to diffraction scintillation on inhomogeneities with a scale $s_d = \lambda /\theta$, where $\lambda$ is wavelength and $\theta$ is angular size of the scattering disk (scale of order $10^7$~cm for  $\lambda = 1$~m), refractive scintillation were observed, which are determined by much larger scales $s_{ref} = \theta R$, where $R$ is the distance from the observer to the effectively modulating layer of the medium ($s_{ref} \sim 10^{12}$~cm). 

The temporal scales of diffractive and refractive scintillation are defined as: $t_{dif} = s_d / V$ and $T_{ref} =s_{ref}/V$, where $V$ is  the velocity of movement of the visual beam relative to the turbulent layer. The velocity of the beam can be determined by both the speed of the source and/or the speed of the observer plus the speed of the medium at which the scattering occurs. If the scattering occurs at the close layer (inhomogeneities near solar system at distances of the order of 10 pc), the main one will be the speed of the observer. There is also a number of other effects caused by inhomogeneities of interstellar plasma, which are well known from the analysis of pulsars observations, but we do not consider them in this paper. 

To estimate the typical temporal and frequency scales of diffraction scintillation and the distances to the source, we used a model of the distribution of free electrons in the Galaxy according to the paper of Yao, Manchester \& Wang 2017 (YMW2017,  \citeauthor{Yao2017} (\citeyear{Yao2017})). Their interactive program\footnote{http://www.atnf.csiro.au/research/pulsar/ymw16/} allows to estimate the distance $R_1$ to the source and the time scale of the scattering, $\tau_{sc}$, at the frequency 1 GHz using known galactic coordinates and the dispersion measure. To estimate the frequency scale of diffraction scintillations $f_{dif}$ at the frequency $f_{LPA} = 111$~MHz, we used the relation $\tau_{sc} = 1.16/(2\pi f_{dif})$ and the frequency dependence $\tau_{sc} \sim f^{-4.4}$, valid for the Kolmogorov spectrum of inhomogeneities. For evaluation of $T_{ref}$ and the refractive scintillation modulation index, $m_{ref}$, we used the following relations (\citeauthor{Gupta1993}, \citeyear{Gupta1993}):

\begin{equation}
T_{ref} [days] = \frac{9.8 \times 10^3}{V[km/s]} \times \sqrt{\frac{R_1[kpc]}{f_{diff}[kHz]}}
\end{equation} 

 \begin{equation}
 m_{ref} = 1.08 \times (\frac{f_{dif}}{f})^{0.167}
  \end{equation}

valid for a statistically homogeneous distribution of inhomogeneities and for the Kolmogorov spectrum. $f_{dif}$ is significantly less than the receiver bandwidth (2.5 MHz), therefore, the diffraction scintillation are significantly smoothed by the band and do not influence the amplitude variations from session to session. For evaluation of $T_{ref}$, we used the value of the pulsar velocity of V = 100 km/s. All considered sources, except J0609+16 and J0640+07, are high-latitude objects and the propagation of emission from them is not influenced by plasma inhomogeneities in the spiral arms of the Galaxy. For them, apparently, a statistically homogeneous distribution of plasma along the visual beam is realized.

\subsection{\textbf{Analysis of structure function}}

Structure functions (SF) are usually used to estimate the typical time scales of amplitude variations due to refractive scintillation structure functions are more preferable than correlation functions when there are gaps in data series, as in our case and there are linear trends in the arrays. $SF$ are determined by the relation:

\begin{equation}
SF(k) = 1/(\langle I \rangle ^{2} N(k)) \sum \limits_{i=1}^{M-i} g(i)g(i+k)[(I(i)-I(i+k)]^{2} ,
\end{equation}      
where $\langle I \rangle$ is a mean value of amplitude, $M$ is the length of data, $k$ is lag in days which is changing from 1 to 0.8 $M$, $g(i) = 1$ when amplitude exists for day $i$ and equal 0 otherwise, $N(k) = \sum g(i)g(i+k)$


In order to remove the rapid internal fluctuations of the sources, the amplitudes for each observation session were averaged and then an adjacent averaging (smoothing) was performed for 15 sessions (days/sessions). Thus, in the smoothed data, it is possible to look for characteristic scales of amplitude variation greater than 8 days.  The smoothing interval was chosen so that the width of the ACF for the non-smoothed array, after removing the noise component, was equal to the width of ACF normalized to the same amplitude for the smoothed array. Noise component for shifts ±2, due to uncorrelated variations of the source amplitude, was interpolated between neighboring points. 

\begin{table}
	\centering
	\caption{Characteristic time and frequency scales of diffraction and refractive scintillation in the direction of the studied sources}
	\begin{tabular}{ccccccc}
	\hline
Name &  $R_1$, & $f_{dif}$, & $T_{ref}$, & $m_{ref}$ & m & $T_{exp}$, \\
     &  pc &  kHz &  days &  &  &  days \\
	\hline
 J0139+33 & 1470 & 0.18 & 277 & 0.12 & 0.31 & 40 \\
 J0302+22 & 1021 & 0.25 & 197 & 0.12 & 0.08 & 44 \\
 J0317+13 & 367  & 0.8 & 68 & 0.15 & 0.08 & 25 \\
 J0609+16 & 1629 & 0.001 & 3720 & 0.05 & 0.17 \\
 J0640+07 & 1257 & 0.008 & 1260 & 0.07 & 0.86 \\
 J0659+14 & 159 & 0.6 & 49 & 0.14 & 0.18 & 27 \\
 J1005+30 & 1990 & 0.3 & 23 & 0.13 & 0.71 \\
 J1132+25 & 25000 & 0.14 & 1300 & 0.11 & 0.1 \\
 J1336+33 & 719 & 2.2 & 55 & 0.18 & 0.27 \\
 J1404+11 & 2183 & 0.3 & 274 & 0.13 & 0.12 & 44 \\
 J1502+28 & 1289 & 0.6 & 140 & 0.14 & 0.071 & 35 \\
 J1538+23 & 1311 & 0.5 & 155 & 0.14 & 0.12 & 34 \\
 J1848+15 & 3546 & 0.002 & 4658 & 0.05 & 0.16 & >600 \\
 J2051+12 & 4066 & 0.015 & 1600 & 0.08 & 0.21 \\
 J2105+19 & 2800 & 0.04 & 804 & 0.09 & 0.09 \\
 J2209+22 & 25000 & 0.01 & 4490 & 0.07 & 0.13 \\ 
\hline
\end{tabular}
\end{table}

The pulsar J0302+22 was chosen for this procedure, for which there was the maximum number of days with pulses: $f_{obs} = 0.8$. With a shorter smoothing interval, intrinsic variations affect the ACF, leading to a decrease in its actual width. The structure function was determined on smoothed arrays, taking into account only those values for which there is a signal exceeding the specified threshold. Fig.~5 shows examples of the corresponding structure functions for eight sources in a logarithmic scale. They have a slope and reach a certain saturation level, which is equal to $2m^{2}$. This level is indicated by a straight line in the figure. Modulation index $m$ was determined either by the amplitudes averaged for each month, if the intermittency factor $f_{obs} < 0.5$, or by smoothed amplitudes with a smoothing interval of 15 days for $f_{obs}$ higher or of an order of 0.5. This limit was chosen so that at least half of the sessions had values of amplitudes exceeding the threshold. For sources with a small number of registered pulses, determining $m$ from smoothed arrays would be incorrect. For all sources shown in Fig.~5 (except of J1848+15) $f_{obs} \geq 0.5$. The obtained modulation indices $m$ in most cases differ significantly from the predicted values $m_{ref}$ calculated from Equation~4. The temporal scale ($T_{exp}$) of amplitude variations is determined by the time shift at which the structure function drops by 2 times from this level. 

Table~3 shows the parameters used to obtain the typical time and frequency scales of diffractive and refractive scintillation. Estimates obtained from experimental data, $T_{exp}$, were done only for pulsars with $f_{obs}$ higher of an order of 0.5, in order to exclude sources with a large intervals of absence of pulses. We have made only one exception for J1848+15. For J1848+15, $f_{obs}$ = 0.16, but the days with the presence of pulses are distributed approximately evenly over the entire time interval (Fig.~2). This source has a high value of the dispersion measure ($DM = 77.42$ pc/cm$^{3}$). For it, the structure function does not reach the saturation level, and therefore we are giving only an upper estimate $T_{exp}$. The columns 2 and 3 show the value of the distance to the source $R_1$ and the frequency scale of diffractive scintillation $f_{dif}$, determined using the model YMW16. In columns 4-6, estimates of the time scale $T_{ref}$ of refractive scintillation in days, modulation index $m_{ref}$ (Equations 3, 4),  modulation index $m$, determined by the amplitudes averaged for each month; column 7 shows the experimental estimates of the time scale from $SF$, $T_{exp}$. Table 4 in the 6th column shows the modulation index $m$, determined by monthly variations in the pulse amplitude. It can be seen that the amplitude variations defined by modulation index $m$ are significantly smoothed out compared to the non-averaged data given in Table~2 in column~11, and $m$ for most sources in several times smaller than $m_1$. The model YMW2017 gives the values of the distances to J1132+25 and J2209+22 exceeding the size of the Galaxy. In the model NE2001 (\citeauthor{Cordes2003}, \citeyear{Cordes2003}) R > 30 kpc for 5 sources from our list, which indicates the need to improve these models. 
For sources with $f_{obs}$ $\geq$ 0.5, the received values of $T_{exp}$ are within the range from 25 to 44 days. For them DM < 24 pc/cm$^{3}$ and distances of the order of 1 – 2 kpc according to the model YMW2017. The values obtained from experimental data, $T_{exp}$, are significantly less than the model (expected) values $T_{ref}$ for both models. Obviously, it is impossible to explain all the observed time scales by refractive scintillation on the interstellar medium. Fig.~2 shows that for J0302+22, J0317+13, J1404+11 there are repeated minimum of $SF$ for time shifts of 11 – 12 months, that is, the annual variations of the amplitude. Harmonics corresponding to this periodicity are present both in the power spectra and in the ACF, as noted above.

One more way to find the typical time scales of the variability, may be to determine the number of pairs of structure function, $N_{k}$ (Equation~5), received at each time shift k.  Herewith, $SF(k)$ is determined by the variations of the average amplitude for each day of observation, in which there is a signal exceeding the specified threshold. This approach better identifies possible periodicities in the case of a small number of pulses over the entire observation interval. Fig.~6 shows examples of such dependencies $N_{k}(k)$ for sources J1132+25 and J0609+16. Here you can also see the semi-annual (significantly smaller amplitude) and annual periodicity in the number of pairs for the corresponding time shift. The annual periodicity is also visible in $N_{k}(k)$ for RRATs J2051+12, J2105+19, J2209+22. Plots for all sources are given on our website\footnote{http://prao.ru/online\%20data/onlinedata.html}

\begin{figure}
	\includegraphics[width=\columnwidth]{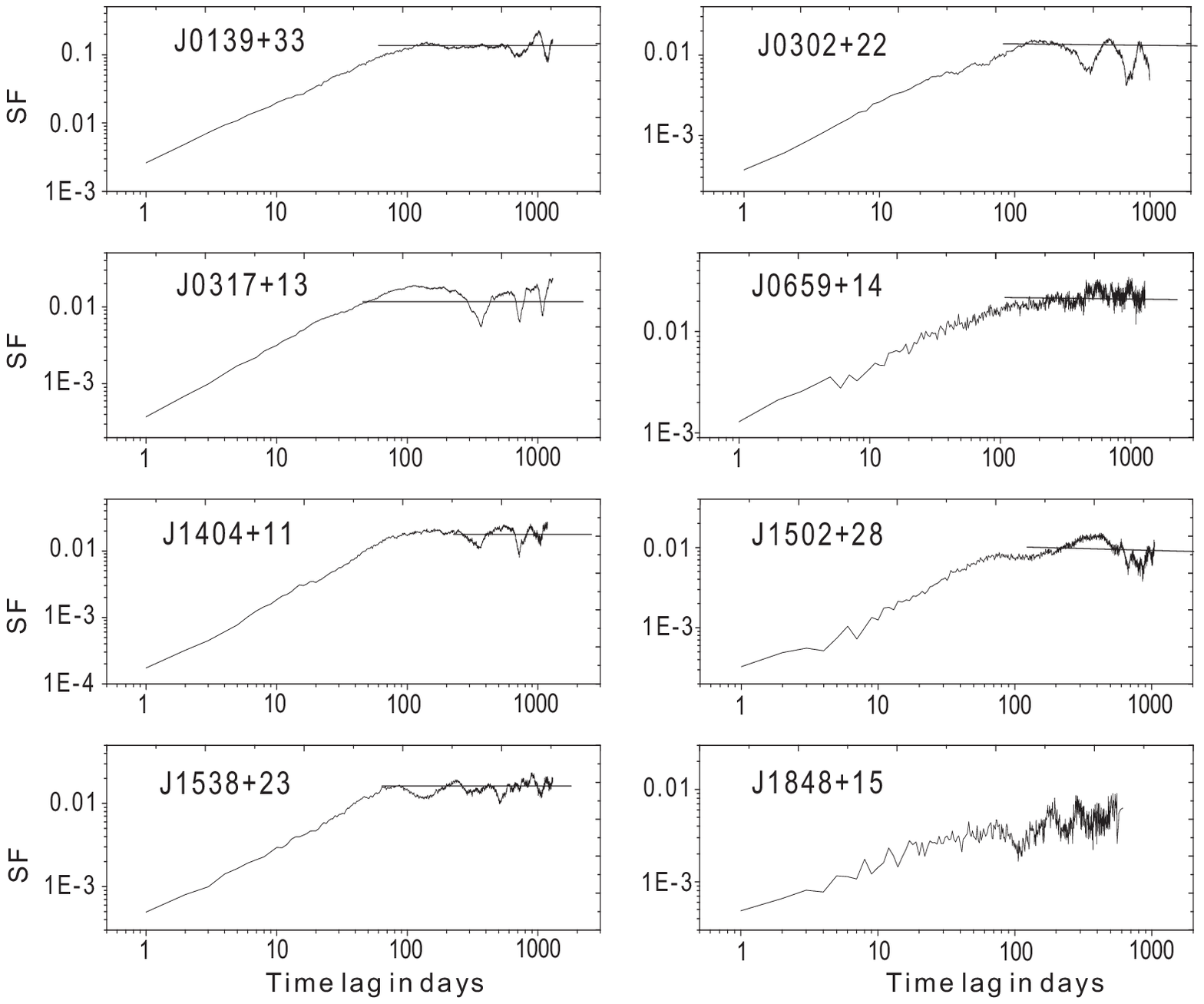}
	\caption{Structure functions for variations in the amplitudes of some sources. The x-axis shows the time shift in days, the y-axis shows the values of the structure function. The straight-line at large time shifts shows the level $2m^{2}$.}
\end{figure}

\begin{figure}
\begin{center}
	\includegraphics[width=0.85\columnwidth]{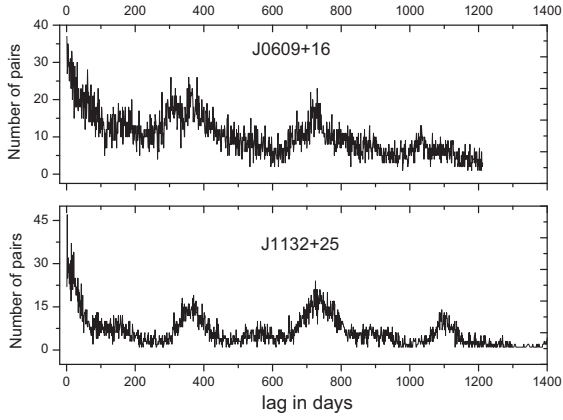}
	\caption{Number of pairs, $N_{k}$ (axis y) in the structure function vs. time shift in days, $k$ (axis x). }
\end{center}
\end{figure}

\subsection{Interplanetary scintillation and annual periodicity}

In the analysis carried out in the section 3.3, the possibility of explaining the temporal variability of the sources due to the passage of emission through the inhomogeneities of the interstellar plasma is considered. It turned out that neither diffraction nor refractive scintillation can explain this variability. In papers  \citeauthor{Gupta1994} (\citeyear{Gupta1994}), \citeauthor{Rickett2001} (\citeyear{Rickett2001}) based on the observations of interstellar pulsar scintillation and a compact extragalactic source, variability was shown on the time scale of a year. This variability is associated with annual variations in the direction of the observer's velocity. However, this mechanism explains only changes in the typical time scale of variations, but it cannot explain changes in the number of observed pulses or their amplitude at time intervals of less than or on the order of two months and with a periodicity of about a year.

The observed periodicity has some peculiarities. First, clearly pronounced narrow (1-2 months) peaks repeated after a year, showing an increase in the number of detected pulses, are not observed for all sources. Second, for different sources, peaks are observed in different months. Suppose that the annual periodicity is associated with the influence of the turbulent plasma of the solar wind, that is, with interplanetary scintillation. In this case, the characteristics of the variability will depend on the relative position of the emission source and the Sun.

Indeed, once a year, any source in the sky gets as close as possible to the Sun. Its elongation (the angle between the directions to the source and to the Sun from the Earth) becomes minimal. For each source, the date of the maximum approach is different and depends on the right ascension and declination of the source. If a compact source passes through the zone of optimal elongations, then its scintillation on the interplanetary plasma is observed.

\begin{figure*}
	\includegraphics[width=0.77\textwidth]{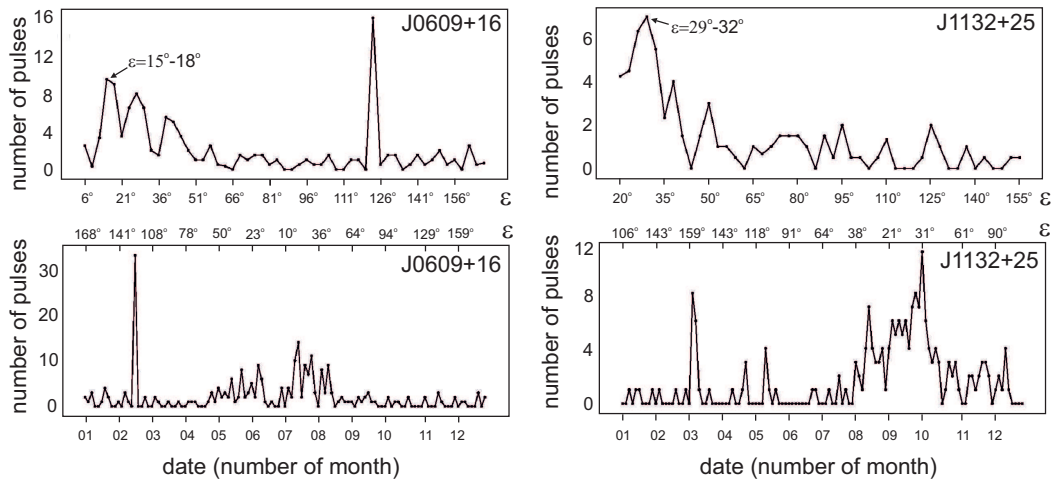}
	\caption{The top panels of the figure show the number of observed pulses (y-axis) vs. the elongation at which they were observed (x-axis). The bottom panels of the figure show the number of observed pulses vs. the observation date. Number of pulses for corresponding dates were averaged over all years of observation.}
\end{figure*}

Let the source under study have many pulses slightly weaker than the threshold for their detection, $S/N=6$. Let the typical duration of these pulses, without taking into account the dispersion measure, not exceed one second, which is comparable to the typical time of interplanetary scintillation. If the interplanetary scintillation are weak, then with the same probability, the $S/N$ of the pulse can increase or decrease. 
Amplified pulses will be detected, weakened pulses will not be detected. However, the weakened pulses would not have been detected in the absence of scintillation anyway. Therefore, the effect associated with interplanetary scintillation of compact radio sources of pulsed emission for pulses with $S/N$ below the detection threshold should lead to an overall increase in the number of recorded pulses. 
On the other hand, let's consider pulses for which the $S/N$ was slightly higher than the detection threshold. For half of them, the observed $S/N$ will become higher, but half of the pulses will disappear, since their $S/N$ will become lower than the detection threshold. Thus, the overall picture can be complex and depend both on the total number of recorded pulses and on their distribution by amplitudes. 
In any case, if a spike is observed in dependence of a number of recorded pulses vs. time, then the number of detected pulses should exceed the number of lost pulses for this source. 

The regime of interplanetary scintillation in the case of a power-law turbulence spectrum is determined by a single parameter: by the value of the $SF$ of the wave phase fluctuations $D_S$ on the Fresnel scale $U_{0}$ = $D_S$ ($z/k$), here $z$ is the distance from the observer to the effectively modulating layer, $k$ is a wave number  (\citeauthor{Prokhorov1975}, \citeyear{Prokhorov1975}, \citeauthor{Vlasov1979}, \citeyear{Vlasov1979}). In the case of $U_{0} < 1$ scintillation is weak and has a diffraction character, in this regime, decreases and increases in intensity are approximately equally likely. In case of $U_{0} \gg 1$ the scintillation is saturated. Similar to interstellar scintillation, intensity fluctuations are a combination of fast diffraction and slow refractive scintillation, and the scintillation index of a point source is close to 1. In the saturation regime, decreases and increases in intensity can also be considered approximately equally probable, but their amplitude is comparable with the average intensity. In case of values $U_{0} \lesssim 1$ a focusing regime is realized (\citeauthor{Prokhorov1975}, \citeyear{Prokhorov1975}), in which the scintillation index of the point source is higher than 1. It can be expected that in the focusing regime, sharp amplifications of the observed emission flux associated with the influence of caustics will be observed. For the frequency of 111 MHz, the transition from weak to saturated scintillation occurs at heliocentric elongations $\epsilon_{m} \approx 25^{\circ}$. The value of $\epsilon_{m}$ depends on the wavelength, $sin(\epsilon_{m}) \sim \lambda$, and it can change from year to year due to variations in the global structure of the solar wind in the solar activity cycle 
(\citeauthor{Manoharan2012}, \citeyear{Manoharan2012}). 

Visible angular dimensions of pulsars $\theta_{0}$ are determined by scattering in interstellar plasma, the typical value of the scattering angle $\theta_{0}$ at our frequency of observations is $\theta_{0}$ $\approx$ 0.1 arcsec (\citeauthor{Artyukh1989}, \citeyear{Artyukh1989}). This value depends on the distance to the pulsar, the galactic coordinates and the wavelength, $\theta_{0}$ $\sim$ $\lambda$ $^{2}$   . For the frequency of 111 MHz, the typical value of $\theta_{0}$ is significantly lower than the angular size of the Fresnel zone, $\theta_{F}\approx$ 0.3 arcsec (\citeauthor{Chashei2021}, \citeyear{Chashei2021}), therefore, the pulsars observed in this paper can be considered as point objects for interplanetary scintillation.

The number of days when the source is in the zone of optimal elongations ($\epsilon = 20-40^{\circ}$), where there is a maximum on the dependence of the scintillation index vs. elongation, usually does not exceed 2-3 months per year, and for some sources, the zone of optimal elongations may be unattainable. On Fig.~2 for RRAT J0609+16 and J1132+25, elongations of $20^{\circ}-40^{\circ}$ are highlighted by segments. It can be seen that the main part of the peaks of short duration coincides with the zone of optimal elongations. 

To illustrate the dependence of the number of observed pulses vs. elongation $N(\epsilon)$, we had chosen pulsars J0609+16 and J1132+25. For both sources, the annual periodicity is visible in Fig.~2. For J0609+16, the minimum elongations reach 6$^{\circ}$, and for J1132+25 the minimum elongations reach 20$^{\circ}$. The total number of observed pulses is small, however, since the elongations reach approximately the same values every year on the same dates, it is possible to average the number of pulses by dates for each year. For additional smoothing, we also performed averaging of $N(\epsilon)$ for elongations with the step of 3$^{\circ}$. Fig.~7 shows the dependencies of $N(\epsilon)$ (top row) and the number of observed pulses vs. a date (bottom row). Here, a preliminary averaging of the number of pulses for three days was carried out.

Fig.~8 shows that the peak of the observed number of pulses for J0609+16 is reached at the elongation $18^{\circ}$, and for J1132+25 is reached at the elongation $30^{\circ}$. As for the dates, these maxima fell into the middle of July for J0609+16 and for the early October for J1132+25. Thus, Fig.~8 clearly demonstrates that there is a dependence of $N(\epsilon)$ with a pronounced maximum close to the zones of optimal elongations. By appearance, the dependence of $N(\epsilon)$ is similar to the dependence of the scintillation index vs. the elongation $m(\epsilon)$, which is well known to researchers using the method of interplanetary scintillation. Elongations at which maxima in $N(\epsilon)$ were observed are close to the elongations at which the focusing regime was expected. In addition to sources J0609+16 and J1132+25, clearly defined maxima in $N(\epsilon)$ dependencies and the number of pulses vs. a date were observed for the sources J0659+14, J2051+12, J2105+19, J2209+22. The maximum in $N(\epsilon)$ dependencies for these sources is close to $\epsilon = 30^{\circ}$. The number of observed pulses at optimal elongations is increasing from 2 to 10 times.

We assume that the annual periodicity of the variability in the number of pulses and the amplitude of pulsars can be explained by focusing on the inhomogeneities of the solar wind in the strong scintillation regime. This interpretation is supported by the fact that there is no such variability in pulsars that do not sufficiently approach the Sun. A sharp increase in the frequency of the appearance of pulses occurs in a narrow zone of elongations $\epsilon_{m}=15^{\circ}$ – $35^{\circ}$, the position of the center of which approximately coincides with the position of the maximum of the interplanetary scintillation index. The characteristics of the observed annual periodicity of the appearance of pulses allow us to make an estimate of the elongation region in which a significant influence of focusing is possible: $\epsilon_{m}=25^{\circ} \pm 10^{\circ}$. 

\section{Discussion}

Seven studied objects (J0139+33; J0640+07; J1005+30; J1132+25; J1336+33; J1502+28; J2105+19), out of the sixteen, are similar to classic RRATs. Regular (periodic) emission was not detected for them. For these sources, with the exception of J0640+07, the given periods were determined as the greatest common denominator of the observed time intervals between pulses. For J0640+07, it was not possible to obtain a period estimate. The median value of the period determined by the seven RRATs is approximately 3.069~s.

For nine objects for which, in addition to strong dispersed pulses, regular emission is detected (J0302+22; J0317+13; J0609+16; J0659+14; J1404+11; J1538+23; J1848+15; J2051+12; J2209+22) the median value of the period is 1.7769~s.  It turns out that in our sample of sources, the median value of the period for pulsars with detected and undetected regular emission differs by 1.5 times. However, the samples are too small to conclude that this difference is meaningful.

We compared these periods with periods in RRATalog\footnote{http://astro.phys.wvu.edu/rratalog}. There are 122 sources in RRATalog and there are periods for 98 of them. The median value of the period is 1.384~s, which is more than the median value of 0.67~s observed for ordinary pulsars, but less than for RRATs investigated in this paper.

For a number of sources studied by us, their intrinsic  variability on different time scales is visually revealed. If we talk about the time scales of seconds and minutes, then for the sources, from one pulse in 3 minutes (J0640+07, J1005+30) to 38 pulses in 3 minutes for J0609+16 were detected. If we consider the largest time scales, then there is a variability with a time scale from six months to twenty months (J1336+33, J1404+11, J1848+15, J2051+12, J2105+19). These sources have burst activity (strong changing of activity) when the number of recorded pulses for 6 months increases for them from 18 (J1336+33) to 300 times (J1404+11). We believe that they are burst pulsars and their emission may be one of the distinctive features of RRATs. 

There are also long-term trends of an increase in the number of pulses by about 2 times at an interval of 5.5 years for J0139+33 and J0302+22. It is possible that for these RRATs, the trend reflects refractive scintillation on the interstellar medium. The typical time of refractive scintillation according to Table~3 for J0139+33 may be $\sim$ 300 days, and for J0302+22 it may be $\sim$ 200 days.

Burst behavior of RRATs and long-term trends of the increase in a number of observed pulses were previously noted in the decimeter range: 1.4 and 2.2 GHz (\citeauthor{Bhattacharyya2018}, \citeyear{Bhattacharyya2018}, \citeauthor{Palliyaguru2011}, \citeyear{Palliyaguru2011}). In the paper \citeauthor{Bhattacharyya2018} (\citeyear{Bhattacharyya2018}) for RRAT 1913+1330 (Fig.~1 and 2 of this paper) there is an increase in the number of recorded pulses at an interval of about 50 months. In the paper \citeauthor{Palliyaguru2011} (\citeyear{Palliyaguru2011}) 8 RRATs were studied. Although the term ``burst behavior'' was not used by the authors, but Fig.~2 in the article shows an significant increase in the number of recorded pulses per hour from the observation epoch for sources J0847-4316, J1754-30, J1819-1458. Long term trends can be during a few years. We believe that these are the same bursts in the form of an increase in the rate of incoming pulses that we have registered. Apparently, burst behavior is an additional characteristic feature of RRAT. The authors also note that it is impossible to explain all the observed time scales by refractive scintillation on the interstellar medium.

The analysis of the distribution of nullings for sources from our sample showed that, in general, it is quite well described by the exponential law, regardless of whether periodic emission is detected in them or not. The smallest deviations from the exponential function have the sources for which the number of events with short nullings in only of one pulse exceeds 60. Distribution width (the parameter $t_1$) increases with an increase in the total number of zero pulses (the amplitude is less than the detection threshold) over the entire observation interval (Fig.~1). In this case the spread of points from a simple exponential drop also increases: basically, this is an excess of the number of events for long nullings.

As is known, the nulling phenomenon consists in the fact that part of the pulses of the pulsar may disappear (\citeauthor{Backer1970a}, \citeyear{Backer1970a}). The total number of known pulsars with nullings is approximately equal to 150  (\citeauthor{Ng2020}, \citeyear{Ng2020}). The part of missing pulses is within very wide limits and can reach up to 95 $\%$.  The relatively small number of known pulsars with nullings compared to the total number of known pulsars is most likely due to the fact that there are small number of pulsars in the sky whose flux density allows to register their single pulses. 

One of the hypotheses about a nature of RRAT (\citeauthor{Zhang2007}, \citeyear{Zhang2007})  consists in the concept that RRATs are pulsars with extreme nullings. Typical time intervals between recorded pulses of RRATs can be from minutes to hours \citeauthor{McLaughlin2006} (\citeyear{McLaughlin2006}). It is easy to estimate that for RRATs with a period of one second, the share of missing pulses can be in the range from 98$\%$ to 99.99 $\%$. For example, for the one RRAT J0139+33 we studied earlier, the share of missing pulses was determined as 99.58$\%$ (\citeauthor{Brylyakova2021}, \citeyear{Brylyakova2021}). In this paper, the number of processed sessions is increased, and according to Table 2, the share of missing pulses is estimated as 98.8$\%$. For ordinary pulsars, the time scale of nulling can be from one period to many hours and even days (\citeauthor{Wang2007}, \citeyear{Wang2007}).

An interesting hypothesis about the nature of RRAT was expressed in the paper  \citeauthor{Burke-Spolaor2009} (\citeyear{Burke-Spolaor2009}). It was shown in the article that the object J0941-39, discovered as RRAT, appears from time to time as a strong pulsar with a low nulling part. The same behavior is demonstrated by the RRAT J0609+16 we studied, for which out of the 1.5 thousand sessions conducted, there was one with a duration of a minute, when it behaved like a strong pulsar with a low nulling part. It is possible that J0941-39 and J0609+16 are a transitional objects between normal pulsars and RRAT. Since the phase of the average profile, when the object is visible as a pulsar, and the phase of the observed pulses, when the object is visible as an RRAT, coincide, the emission mechanisms must also be related.

There are two main types of internal pulse activity. The first is a very strong one from pulse to pulse variability, the second is nulling. In both cases, for the observer, objects can be detected as RRAT according to the definition of \citeauthor{McLaughlin2006} (\citeyear{McLaughlin2006}). That is, there can be at least two types of RRATs: 1) a regular pulsar $\to$ pulsar with individual missed pulses $\to$ a pulsar with a high number of nullings $\to$ pulsar with observed individual pulses (we call it RRAT); 2) a pulsar whose individual pulses are very strong, and the rest pulses are very weak, and therefore it is detected as RRAT. 

As noted in this section above, half of the objects studied by us fit the first type, and the second half fits the second type. In the paper \citeauthor{Burke-Spolaor2009} (\citeyear{Burke-Spolaor2009}), it was also said that there may be an uncompromising RRAT class that is not included in the types discussed above.

For eight RRATs, the data analysis shows the annual frequency either by the amplitude or by the number of recorded pulses. Since the time intervals of this increase are close to the dates at which the maximum is observed in dependence of the scintillation index vs. the heliocentric elongation, we believe that this periodicity is associated with the interplanetary scintillation. The interplanetary scintillation can significantly increase the number of recorded pulses. Generally speaking, the interplanetary scintillation of compact radio sources is observed at frequencies from gigahertz to tens of megahertz, but they are of particular importance at wavelengths from several meters to a meter. At these wavelengths, the zone of optimal elongations, where noticeable scintillation is observed, overlaps large spatial regions following the movement of the Sun.  

For some RRATs, there is also a weak semi-annual periodicity. In the paper  \citeauthor{Chashei2013} (\citeyear{Chashei2013}) solar activity was studied using the method of interplanetary scintillation. It turned out that scintillation with a typical time scale $\sim$ 1s is also observed in the anti-solar direction, similar to the typical time of interplanetary scintillation. The nature of scintillation in the anti-solar direction was not discussed in the paper. It is possible that a slight increase in the scintillation at time scale $\sim 1$~c at high elongations is due to the influence of inhomogeneities in the electron concentration in the night tail of the Earth's magnetosphere.

In Fig.~7b of the paper \citeauthor{Chashei2013} (\citeyear{Chashei2013}), it can be seen that scintillation in the antisolar direction are much weaker than scintillation at elongations corresponding to the maximum of the scintillation index, and are not visible every year. That is, if a source has a sharp increase in the number of pulses or integral amplitudes at optimal elongations and scintillation in the anti-solar direction is observed for the same source, then when searching for an annual periodicity, a semi-annual periodicity can also be detected. We cannot verify this assumption using our own data, since we cannot estimate the scintillation index of the studied RRATs, neither in the direction of the maximum scintillation, nor in the anti-solar direction, due to too low integral densities of the source fluxes and small number of pulses.

The practical conclusion for researchers engaged in the search and study of RRATs in the meter wavelength range is as follows: 1) the search for new RRATs should be better carried out in the zone where dependence of the scintillation index vs. the elongation has the maximum; 2) the effect of a possible increase in the number of observed pulses should be taken into account when constructing various kinds of dependencies. For example, when developing a dependence of the number of pulses vs. the observed $S/N$, when searching for the intrinsic variability of sources, and others.

We do not consider in detail the features of the detected annual variations in the observed number of pulses vs. the source elongation $N(\epsilon_{m})$. There may be other effects that increase or decrease the number of detected pulses. In particular, there should be an effect associated with an elevation of the source above the galactic plane. The closer to the plane of the Galaxy, the higher will be the scattering in ISM, and the moment will come when the source will become extended (not scintillation), and the maximum in the dependence for $N(\epsilon_{m})$ will disappear. The effect also depends on the wavelength of the observations, since the scattering angle depends on the wavelength as $\lambda^2$. That is, the longer the $\lambda$, the stronger the effect. Another possible effect is related to the angular resolution of the radio telescope. It may turn out that several compact (scintillating) sources will fall into the antenna receiving beam, and then, in addition to the pulsar's own variations can be influence from the sources of confusion.

\section{Conclusion}

In summary, we detect the annual and semi-annual periodicity, observed as a noticeable increase in the number of sessions with detected pulses, the variability on typical time scales from months to a year and a half, the long-term trends in the change of the number of observed pulses, nullings of different durations. 

 The annual and semi-annual periodicity cannot be explained by the effects of scintillation on interstellar plasma and it is possibly associated with scintillation on interplanetary plasma. For two RRATs (J0139+33 and J0302+22), there are pronounced trends in the change of the number of detected pulses by  over the entire observation interval.

For some RRATs (J1336+33, J1404+11, J1848+15, J2051+12), their intrinsic variability is observed as increase in the number of  pulses or their amplitude by many tens of times at intervals from months to years. The RRAT J0609+16 has unusual behavior when strong activity during 1 minute changes rapidly to weak one. 
It is shown that the distribution of the times of the nulling durations is exponential. 

\section*{Data availability}
The raw data underlying this article will be shared on reasonable request to the corresponding author. The tables with $S/N$ of pulses and additional figures  are in http://prao.ru/online\%20data/onlinedata.html

\section*{Acknowledgements}

The authors are grateful to the technical team of LPA LPI for assuring observations in the monitoring regime. Special thanks to L.B. Potapova for a help with preparing of pictures.


\end{document}